# Electric-field Induced Reversible Switching

## of the Magnetic Easy-axis in Co/BiFeO₃/SrRuO₃/SrTiO₃

## Heterostructures

T.R. Gao[1], X.H. Zhang[1,2], W. Ratcliff II[3], S. Maruyama[1], M. Murakami[1], V. Anbusathaiah[1], Z. Yamani[4], P.J. Chen[2], R. Shull[2], R. Ramesh[5], and I. Takeuchi[1,*]

[1]Department of Materials Science and Engineering, University of Maryland, College Park, Maryland, 20742, USA

[2]Materials Science and Engineering Division, National Institute of Standards and Technology, Gaithersburg, Maryland 20899, USA

[3]NIST Center for Neutron Research, National Institute of Standards and Technology, Gaithersburg, Maryland 20899, USA

[4]Canadian Neutron Beam Centre, National Research Council, Chalk River Laboratories, Chalk River, Ontario, Canada K0J 1J0

[5]Department of Materials Science and Engineering, University of California, Berkeley, California 94720, USA

*takeuchi@umd.edu

Electric-field (E-field) control of magnetism enabled by multiferroics has the potential to revolutionize the landscape of present memory devices plagued with high energy dissipation. To date, this E-field controlled multiferroic scheme at room temperature has only been demonstrated using BiFeO$_3$ (BFO) films grown on DyScO$_3$ (refs 1 and 2), a unique and expensive substrate, which gives rise to a particular ferroelectric domain pattern in BFO. Here, we demonstrate reversible E-field-induced switching of the magnetic state of the Co layer in Co/BFO (001) thin film heterostructures fabricated on SrTiO$_3$ substrates. The angular dependence of the coercivity and the remanent magnetization of the Co layer indicates that its easy axis reversibly switches by 45° back and forth between the (100) and the (110) crystallographic directions of SrTiO$_3$ as a result of alternating application of positive and negative voltage pulses on BFO. The coercivity of the Co layer exhibits a hysteretic behavior between two states as a function of voltage. To explain the observation, we have also measured the exact canting angle of the antiferromagnetic G-type domain in BFO films for the first time using neutron diffraction. These results suggest a pathway to integrating BFO-based devices on Si wafers for implementing low power consumption and non-volatile magnetoelectronic devices.

E-field control of magnetism by integrating a multiferroic material is widely being pursued because of its potential in spintronic applications with reduced energy dissipation due to the intrinsic coupling between magnetic and ferroelectric (FE) order parameters in multiferroic materials[1-9]. As the only known single-phase multiferroic material that exhibits both ferroelectric (FE) and antiferromagnetic (AFM) order parameters at room temperature[10,11], BFO has attracted great attention. It has been demonstrated that the AFM state of BFO films can be reversibly changed with an E-field due to the coupling between its FE polarization and AFM spins[12,13]. The manipulation of the AFM state in BFO, in turn, can be used to influence the properties of an adjacent ferromagnetic (FM) layer through exchange coupling at the interface between the AFM BFO layer and the FM layer[14-17].

Taking advantage of the magnetoelectric coupling and exchange coupling, Heron et al. have demonstrated a 180° reversible switching of the magnetization of CoFe in CoFe/BFO heterostructures as the E-field is applied horizontally or vertically through the BFO films[1,2]. Their findings have shown that the switching of magnetization is directly associated with the FE domain structure of BFO films deposited on $DyScO_3$ (DSO) substrates which only display one set of FE stripe domains. In the present work, by using BFO films grown on $SrTiO_3$ (STO) substrates, we demonstrate reversible 45° switching of the magnetic easy axis in the Co layer exchange coupled to BFO by applying alternating vertical voltage pulses. We show that concomitant with the switching, the coercivity of the Co layer is reversibly tunable between two states. A key difference between the reported work on BFO films fabricated on DSO substrates and our work on BFO films fabricated on STO substrates is the difference in the switching behavior of the ferroelectric polarization, leading to the different switching angle of the ferromagnetic magnetization of the Co layer. This new switching mechanism in the present work was previously overlooked, and because it does not rely on a unique ferroelectric domain structure of BFO on DSO, it is more generic, and it can be extended to STO grown on $Si$[18-20]. Thus, the present work is an important step toward epitaxial BFO-based E-field tunable spin-valve devices integrated on Si.

Magnetic hysteresis loops were measured on 100 µm × 100 µm and 200 µm × 200 µm Co pads at room temperature using a longitudinal magneto-optical Kerr effect (LMOKE) setup. In order to enhance the signal/noise ratio of the LMOKE response from each of the patterned pads, hysteresis loops were averaged over 10-15 measurements. Figure 1 shows typical hysteresis loops of a 100 µm × 100 µm Co pad on an 80 nm thick BFO film with the magnetic field applied along the STO (100) direction (black loop) and the STO (010) direction (red loop). The magnetic in-plane coercivity ($H_C$) is approximately 11 mT and 7.5 mT, and the remanence ratio ($M_R/M_S$) is 0.85 and 0.6 when the field is applied along (100) and (010) directions, respectively. The $H_C$ of the Co layer is larger than that of Co films directly deposited on STO substrates ($\approx$ 1.5 mT), indicating that a robust exchange coupling is established between the FM Co and the AFM BFO layer. However, the exchange bias field ($H_E$), i.e. the shift of the hysteresis loop along the field axis, is quite small with an upper limit of 0.3 mT. The small exchange field could be due to the small antiferromagnetic anisotropy of BFO films[21,22], and this is consistent with previous reports of the exchange bias field of CoFe layers on BFO films where BFO heterostructures with stripe FE domain structures show relatively small exchange bias field[14].

In order to determine the direction of the easy axis of the Co layer and investigate the E-field effect on its magnetization state, the angular dependence of the hysteresis loop was measured using LMOKE relative to the (100) or the (010) direction of the BFO/STO in the as-deposited state and after the voltage pulses were applied. The voltage pulses, $\pm$ 3.5 V corresponding to $\pm$ 400 kV/cm were applied vertically to the BFO layer between a bottom SRO electrode and a Co pad, and after each voltage pulse, the angle-dependent MOKE hysteresis loop was measured. From the measured hysteresis loops, the $H_C$ and the $M_R/M_S$ can be mapped at angles relative to the (100) or the (010) direction of STO. A schematic view of the device is shown in the inset of Figure 1.

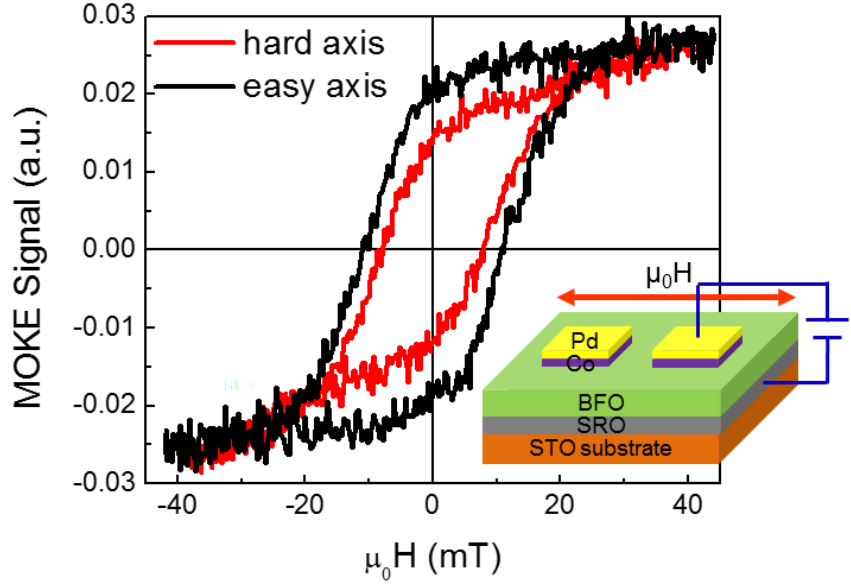

Figure 1 Typical longitudinal magnetooptical Kerr effect (LMOKE) hysteresis loops of a Co pad in Co (5 nm)/BiFeO$_3$ (80 nm) film heterostructure at hard axis (red) and easy axis (black). The inset shows the device schematic for LMOKE measurements. The magnetic field is applied in the in-plane directions (red arrows) during the MOKE measurement. The typical pad size is 100 μm × 100 μm. Note: 1 mT = 10 Oe.

Figure 2 shows the angular dependence of H$_C$ and M$_R$/M$_S$ in the as-deposited state ((a), (e)), after the first + 400 kV/cm poling ((b), (f)), after the subsequent - 400 kV/cm ((c), (g)) and after the second + 400 kV/cm poling ((d), (h)). ± 400 kV/cm corresponds to ± 3.5 V, and for each poling, the voltage is applied for approximately 10 seconds. The H$_C$ and the M$_R$/M$_S$ both reach maxima at 0° (and 180°) and minima at 90° (and 270°) in the as-grown state ((a) and (e)), indicating that the easy and the hard axes of the Co pad are along the (100) and the (010) crystallographic axes of the STO layer (i.e., 0° and 90° in the polar curves), respectively. (We note that because the (100) direction and the (010) direction here are equivalent, for a given device, we denote the initial easy axis direction as the (100) direction.) The polar curves have the typical shapes of angular dependence of the $H_C$ and the $M_R/M_S$ in a conventional ferromagnet/antiferromagnet exchange-biased system[23,24].

After a positive voltage (+ 400 kV/cm) pulse was applied to the same pad, the $H_C$ and the $M_R/M_S$ now exhibit their maxima and minima at 45° and 135°, respectively (Fig. 2(b) and (f)). The results suggest that after the E-field poling, the easy and hard axes are now located at 45° and 135°, respectively. The easy axis now is along the (110) or (-110) directions. As shown in Fig. 2(c) and (g), after applying a - 400 kV/cm E-field pulse on the same sample pad, the easy and the hard axes switch back to 0° and 90°, respectively, which are the starting crystallographic (100) and (010) directions of the STO film. Fig. 2(d) and (h) show the angular dependence of $H_C$ and $M_R/M_S$ after the second positive E-field pulse (+ 400 kV/cm) was applied. The easy axis has once again rotated 45°.

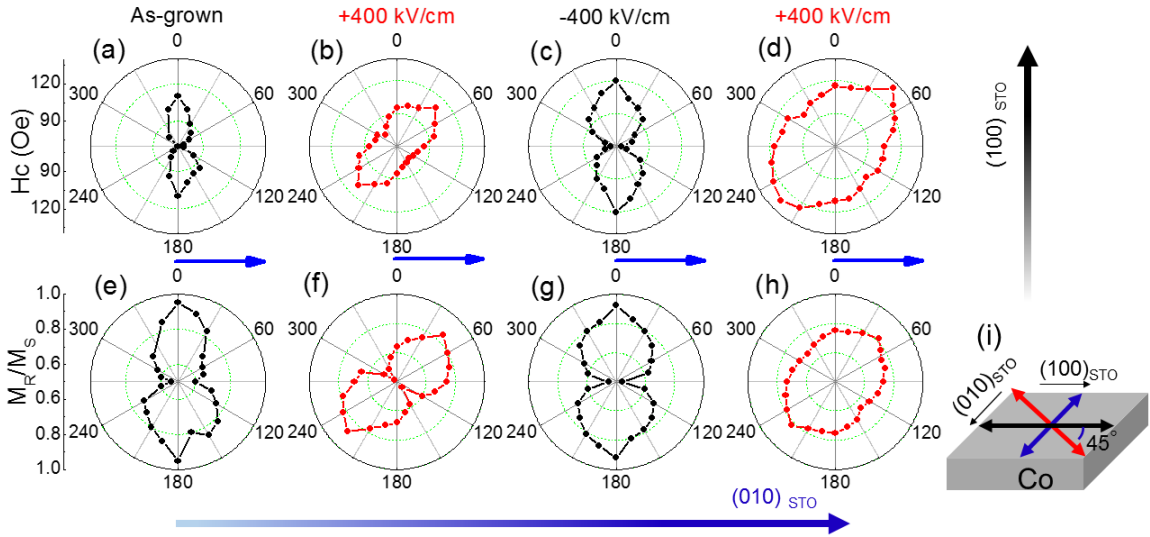

Figure 2 Polar curves of the angular dependent magnetic coercivity and Mr/Ms of a Co layer pad in Co (5 nm)/BFO (80 nm) heterostructure in as-grown state ((a) (e)), after the first + 400 kV/cm poling ((b) (f)), after the subsequent - 400 kV/cm poling ((c) (g)), and after the second + 400 kV/cm poling ((d) (h)). 0 and 90 degrees indicate the (100) and (010) directions of STO substrate (long blue and black arrows), respectively;

(i) schematic of 45° switching back and forth of the Co easy axis between (100) and (110) directions. The black and red arrows in (i) represent the easy axis directions of the Co layer in as-grown and (the initial) poled states, respectively.

The results here clearly demonstrate that the easy axis of the Co layer reversibly switches back and forth with applied polarity-alternating E-field pulses. This indicates that there is reversible change in the AFM domains of the BFO film, which determines the exchange bias and its direction in the Co layer. The AFM domains are in turn directly associated with the FE polarization of the BFO film[13].

It has been explained theoretically that in strained BFO films with stripe domains, there are G-type AFM domains with local AFM moments oriented in the (111) planes with canted moments in the (11-2) direction[2,25]. It is the projection of this canted moment to the surface of the film which leads to exchange bias. There have previously been numerous experimental reports on investigation of the magnetic structure of BFO films by neutron diffraction[26-29]. The general consensus is that in strained films much thinner than 500 nm, there are G-type AFM domains, while when the films get thicker, cycloids begin to appear. We note that due to the subtle differences in film-growth conditions among different groups and the subsequent residual strain states of the films, the "cross-over" thickness (G-type to cycloid) differs from report to report. Most recently, Bertinshaw et al. have showed that they can grow strain-relaxed BFO films down to 100 nm, and have observed cycloids at this thickness[27]. As we discuss below, our films used in the present study (80 – 200 nm thick) exhibit clear stripe domains, and they thus consist of G-type AFM domains, which are needed to give rise to the clear exchange-bias in the Co film deposited above.

Even though this canted moment geometry is consistent with the observed exchange bias behavior[22], the details of the canting has never been experimentally studied. To this end, we have carried out neutron diffraction measurements in the (HHL) scattering plane in a BFO film (200 nm) on the C5 triple axis spectrometer at Chalk River Laboratories in magnetic field at H ≈ 0 (except for a small guide field less than 3 mT).

Figure 3 shows reciprocal space maps of a 200 nm thick BFO film revealed by neutron diffraction where only one reflection is present, which indicates that the magnetic structure of BFO film is G-type, rather than cycloids as seen in a previous study[26]. In the previous study, the period of the cycloid was found to be longer. However, with our resolution (indicated in red on the figure), we would have still seen the peak splitting. Thus, we are confident that these films (80 – 200 nm thick) consist of G-type domains, rather than magnetic cycloids.

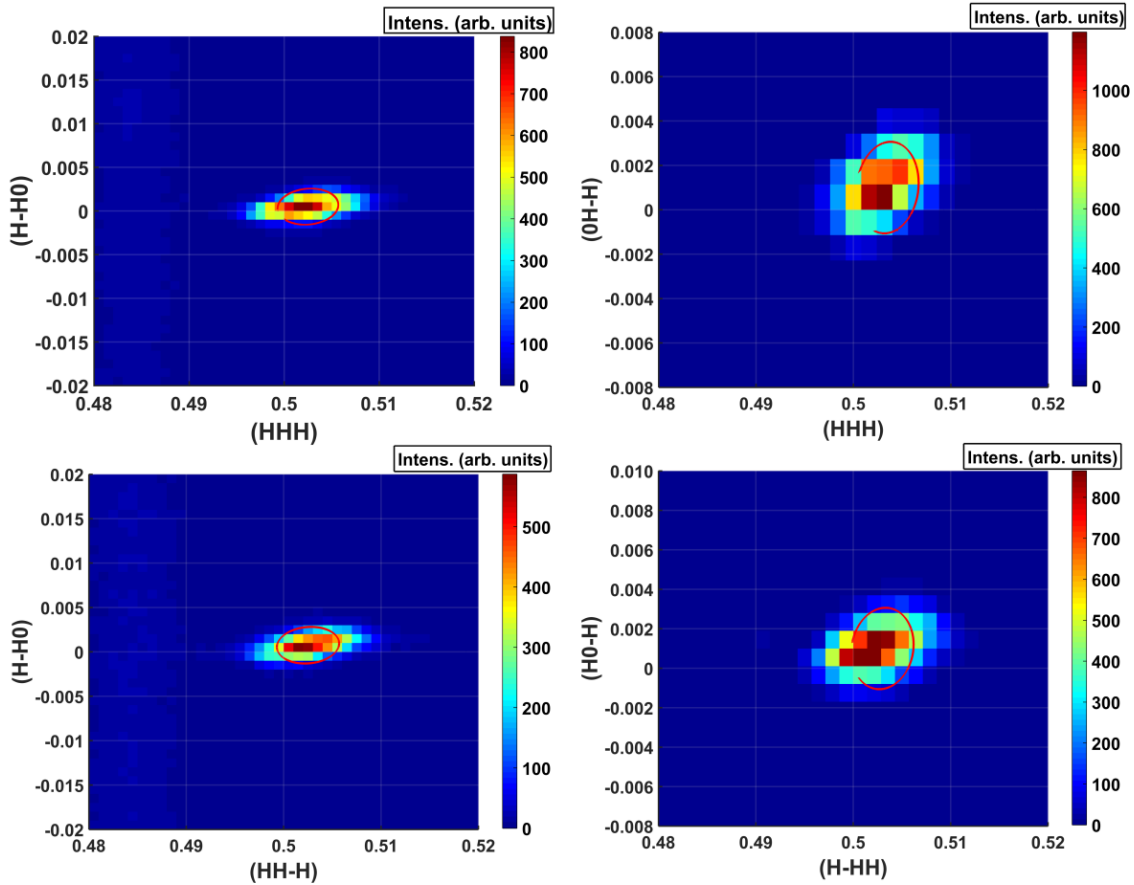

Figure 3 Reciprocal space maps of a 200 nm thick BFO film measured in different scattering planes using neutron diffraction. Red lines represent the instrumental resolution.

To precisely determine the direction of magnetic moments of the 200 nm thick BFO film, we turned to polarized neutron diffraction measurements in the HHL zone and measured the (.5 .5 .5) and (-.5 -.5 -.5) magnetic reflections. Several nuclear reflections from the substrate were measured to determine the flipping ratio and the uniformity of the spin transport, and flipping ratios of $\approx 15$ were achieved. As there is a 3-fold axis about the (111) axis for rhombohedral systems, we expect magnetic domains to be present. Given equally populated magnetic domains such as in BFO, we are only sensitive to the magnitude of the magnetic moment and its angle with respect to the (111) axis. We are not able to determine the moment on an absolute scale due to our inability to measure nuclear reflections from the film, which will be shadowed by reflections from the substrate. From our fits, we find the angle of the magnetic moment with respect to the (111) is 83.9(23) degrees ($\approx 84°$) which is consistent with the magnetic easy plane suggested by Ederer et al.[25] In the fitting, we have corrected for non-uniform spin transport and for the finite flipping ratio. These results are presented in the supplementary materials. We note that previous studies have shown that the magnetic structure of these materials is very sensitive to strain, growth direction, and synthesis conditions[28,29].

We have investigated the ferroelectric domain structures of the BFO films on STO substrates and their switching modes by piezoresponse force microscopy (PFM). The details of the PFM characterization technique is described elsewhere[13]. Briefly, both out-of-plane and in-plane components of the FE polarization are simultaneously acquired from the PFM images obtained using two lock-in amplifiers. Figure 4(a) shows the in-plane PFM images of a 80 nm thick BFO film with the PFM cantilever oriented along the (110) direction in the as grown state and after 3.5 V poling. Here, the voltage was vertically applied between two electrodes, i.e. an SRO layer and a Pd layer (deposited above BFO for this study). To identify the polarization direction in each ferroelectric domain and their switching modes, we have taken the in-plane and the out-of-plane PFM images (not shown here) along two orthogonal (110) directions of the STO substrate. In the as-grown state, the stripe domains are clearly observed, and two colors are seen, which correspond to two possible polarization variants[30]. Note that Fig. 4(a) only shows one of the two sets of domains in the as-grown state. The other set of domains would be appear orthogonal to the one seen here. Based on both in-plane and out-of-plane PFM images, we are able to deduce these two variants to be pointing in (-1-11) and (-111) directions. The projections of the two variants on the (001) plane marked with blue arrows in (a). After the sample is poled at 3.5 V (400 kV/cm), the domains are now split into two sets of stripe domains orthogonal to each other: the shape/direction of some of the domains has rotated $90^\circ$ with respect to the initial shape/direction. This change is allowed because on BFO films grown on STO the two sets of orthogonal domains are equally likely. Concomitantly, the domain width has also increased after E-field poling, indicating that some initial domains with different polarization directions have now merged together into a single polarization direction. Within a fixed PFM region (5 μm × 5 μm), we are able to distinguish the different switching modes (71°, 109°, and 180° switching) after E-field poling and to evaluate the switching mode fractions (Fig. 4(c)). It can be seen that 54% and 40% of the area have undergone 71° and 109° switching, respectively, and 180° switching has taken place in less than 10% of the area. These observations are consistent with the 45° switching mechanism of the magnetic easy axis of the Co layer deposited on top of the BFO layers as discussed in the next section.

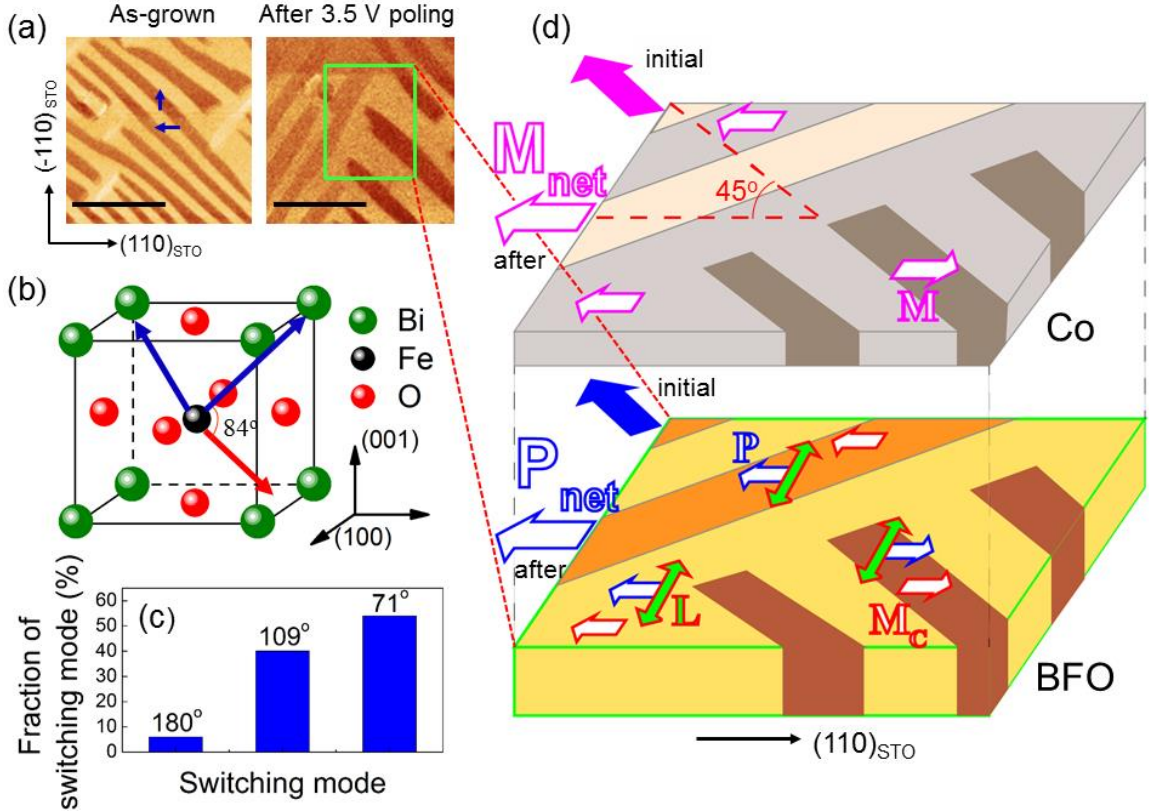

Figure 4 (a) In plane PFM images of 80 nm thick BFO film before and after E-field poling (same area); here the PFM cantilever is pointing along (110) direction of STO substrate during PFM measurements. The images were cropped from 5 μm ×5 μm scan area images. The scale bars are 1 μm. (b) Unit cell of BFO; the blue arrows represent two FE polarization variants, and the red arrow represents the canted magnetic moments with respect to one of the polarization variants. The canted angle of the moment relative to the polarization pointing to (-111) is 84°. (c) Fractions of different switching modes observed here. (d) Schematics of magnetoelectric coupling between BFO and Co. The FE domain configuration of BFO in (d) is a representation of PFM image inside the green square in (a). The small unfilled blue arrows represent the projections P of switched FE polarizations on the (001) plane of BFO after poling; the large solid and unfilled blue arrows represent the net polarization projections on the (001) plane $P_{net}$ before and after E-field poling, respectively; the unfilled red arrows represent the projection of canted moments $M_C$ after switching; the solid green arrows represent the antiferromagnetic axis L of BFO; the small unfilled pink arrows represent the magnetic moments of Co exchange-coupled to the projection of the canted moments in BFO after switching;

the large solid and large unfilled pink arrows represent the easy axis direction of Co before and after polarization switching, respectively: the easy axis direction has switched by 45°.

Fig. 4(b) shows the schematic of two initial polarizations (-1-11) and (-111) directions (blue arrows). The angle between the FE polarization and the canted magnetic moment is 84° as determined by neutron diffraction, and it is shown for one of the FE polarization arrows. As discussed above, it is the projection of the canted G-type AFM moments on the (001) plane which generates the exchange bias on adjacent FM layers, and the projection direction is parallel to the projection of the polarization on the plane. Fig. 4(d) shows the schematic of the coupling relationship between the polarization in the BFO layer and the coupled magnetic moment in the Co layer. Because we initially have equal populations of (-1-11) and (-111) domains, the net polarization projection points to the (-100) direction, as represented by the large solid blue arrow. The ferroelectric polarization and the canted moments are directly linked together with 84° via the antiferromagnetic axis L (green arrows in Fig. 4(d))[1,2,12], and the in-plane projection of the moment is always parallel to the in-plane projection of the polarization[2]. The end result is that the projection of the canted AFM moments on the (001) plane initially points to the (-100) direction as an average over this region. Thus, the Co layer always has the initial easy axis in this direction (-100) represented by large solid pink arrow in the Co layer in Fig. 4(d). We note that we do not apply biasing magnetic field during deposition of the Co-layer on BFO and thus the easy axis direction is entirely dictated by the distribution of ferroelectric local domains in BFO. As deduced from PFM images, most of the domains undergo 71° and 109° switching (Fig. 4(c)), and the two starting polarizations now point either both in the (-1-1-1) direction or in the (-1-1-1) and (11-1) directions after switching. Therefore the net polarization on the (001) plane switches 45º or 135º and points to the (-1-10) (large unfilled blue arrow) or the (110) direction. Of the two, the 135º switching is unlikely because it forms energetically unfavorable tail-to-tail domain configuration as discussed in the next section.

Because of the bi-axial strain applied to BFO by STO, the other set of FE domains orthogonal to the region shown in the initial state of Fig. 4(a) is likely present on other parts of the same BFO film. The polarizations in these FE domains would naturally switch with the same fractional percentage modes as the polarizations shown in Fig. 4(a). The net polarization directions of a local region before and after E-field poling depends on which set of domains is dominant, which can vary from sample to sample and the film growth conditions[30], and locally it is always 45º rotated with respect to the net polarization before poling as described above.

Let us examine the further details of the switching to see how we end up with a 45° angle but not any other angles. All possible combinations of the switching of the easy axis of the Co layer as a result of switching of two initial polarization directions are shown in Fig. 5. The red arrows depict the average projection of the canted moments of BFO on (001) plane. Because the moments are tied to the polarizations, the in-plane directions depicted here simultaneously represent both the polarization projection and the canted-moment projection, which determines the local direction of the unidirectional exchange bias the Co layer experiences. The initial two adjacent FE polarizations of BFO point to (-1-11) and (-111) directions (Fig. 4(b)), corresponding to coexisting variants within one set of domains, and their projections on the (001) plane of the average polarization is parallel to the net magnetic moments of the Co layer indicated by the solid pink arrow in Fig. 4(d). After poling with a 3.5 V pulse, each polarization can undergo three possible switching: 71°, 109° or 180°. Given the initial projection directions as shown in Fig. 4(b) then, this results in six possible combinations of relative domain configurations for the adjacent domains after switching: namely the two switching are by: 71°/71°, 109°/109°, 180°/180°, 71°/109° (or equivalently 109°/71°), 71°/180° (or equivalently 180°/71°), and 109°/180° (or equivalently 180°/109°) where the first number is the switching angle of the left domain in Fig. 5 (a)-(f), and the second number after / is the switching angle of the right domain in Fig. 5(a)-(f). It has been reported that direct 180° switching in strained BFO films fabricated on STO is highly unfavorable energetically and that the two-step 180° switching is likely observed only in small-sized devices (e.g., 3 μm × 3 μm)[31]. Thus, for our large devices (100 μm × 100 μm), we eliminate three of the six configurations which involve 180° switching in Fig. 5 which are (c), (e), and (f). There are thus three remaining possible switching configurations in Fig. 5 ((a), (b), and (d)). Of the three, only the 71°/109° switching case results in head-to-tail polarization domain boundary condition, which is energetically favorable over other configurations which result in head-to-head or tail-to-tail boundary configurations which are electrostatically unfavorable. Thus, out of all possible switching configurations, the 71°/109° case is the predominant one. This is also consistent with the fact that we observed more regions with 71° and 109° switching than ones with 180° switching (Fig. 4(c)). As seen in Fig. 5(d), this configuration leads to average direction of the projected moment at 45° from the (100) direction, resulting in exchange bias with the easy axis now rotated 45°. It is known that the same but reversed path of the polarization rotation is the favorable one upon application of a reverse field[2]. This results in the domains returning to the original domain configurations with the second pulse of – 3.5 V, thus leading to the reversible switching of the exchange-biased easy axis of the Co-layer between (100) and (110) directions as observed here.

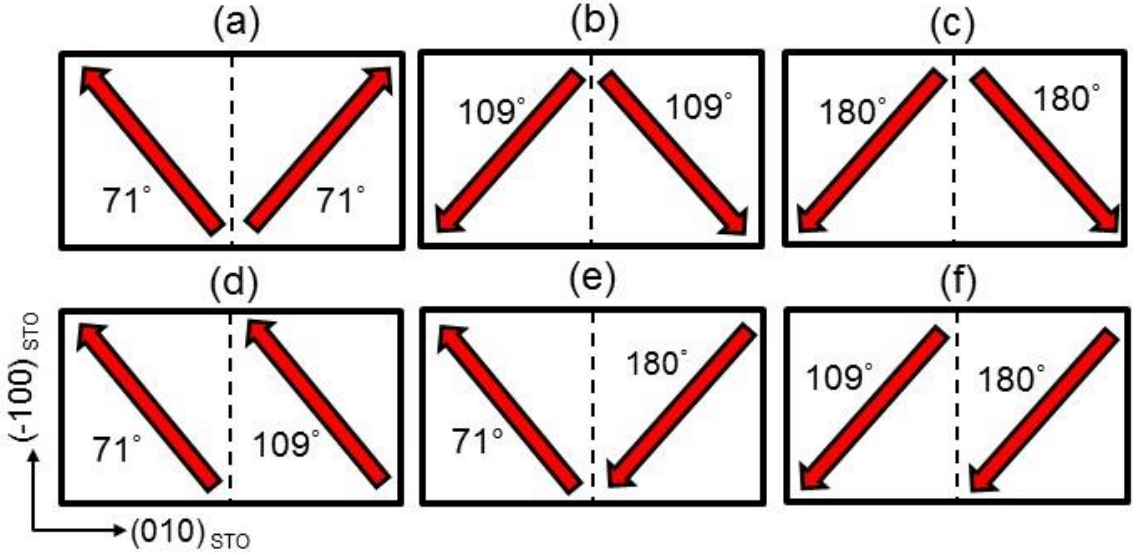

Figure 5. Six possible switching combinations of the canted moments in the BFO layer after the first positive E-field poling. The red arrows represent the projection of moments on the (001) plane after switching. The inset numbers indicate the switching angles of polarization, which results in the corresponding moments switching, and the dash lines is the FE domain boundary. Note that in the cases (d) and (f), the domains would merge to be one.

The difference between the deterministic 180° magnetization switching of a CoFe layer in CoFe/BFO/SRO/DSO structures previously reported by Heron et al.[1,2] and the 45° switching in our structures on STO is attributed to two facts. First, the different substrates give rise to different strain states imposed on BFO films, and thus there are different FE domain structures in BFO. Specifically on DSO substrates, only one set of stripe domains are observed in epitaxial BFO[1,2,32], as opposed to the two sets observed in BFO on STO. DSO has an orthorhombic structure with the lattice constant a = 5.54 Å, b = 5.71 Å and c

= 7.89 Å (pseudocubic lattice constant 3.944 Å), whereas the lattice constant of STO is a = b = c = 3.91 Å. The difference in the lattice mismatch between the substrates and BFO (a = b = 3.96 Å, c = 4.02 Å) is such that the symmetry in relative populations of different polarization domains of strained (001) BFO films grown epitaxially on DSO is broken, leading to only one set of domains in BFO on DSO[32,33]. On STO, on the other hand, BFO experiences bi-axial strain, resulting in roughly equal proportion of different polarization domains giving rise to the observed PFM images with two sets of stripe domains[30,33].

Second, it is known that the different device sizes and varying FE domain sizes can lead to different switching behaviors because the stability of 71° switching is associated with the domain size[31]. As reported in Ref. 1 and 31, the deterministic 180° switching in the 3 μm × 3 μm sized device is a two-step switching process consisting of a 71° switching and a 109° switching, where the initial 71° switched state relaxes via a subsequent 109° switching. The relaxation rate of the "intermediated" 71° switched state depends on the local FE domain volume and is inversely proportional to the E-field poled area due to the depolarization from the unswitched area[1,31]. In our 45° switching case, the area size of the Co pad (E-field poled area, see the cartoon in Fig. 1) is 100 × 100 (or 200 × 200) μm[2] which is much larger than the E-field poled area in the device areas reported by Heron et al. and by Baek et al., where the 180° switching was observed[1,31]. Thus, the relaxation rate from 71° switching to 180° switching in the present devices is expected to be much longer, or the relaxation rate is so long that the subsequent switching does not take place. In other words, the 71° switching is stable and the 180° switching (71° followed by 109°) does not take place in our relatively large devices. Therefore, 45° switching of magnetization in the exchange-coupled adjacent FM layer is dominant if the BFO film is grown with bi-axial strained (on STO) which results in two sets of FE domains and if the E-field is applied to a large area (≈ 100 μm region) in BFO film, whereas two-step 180° switching of magnetization is dominant in a smaller E-field poling area (e.g., 500 nm region) in BFO films grown on anisotropic structural substrates such as DSO. We note that the minimum size of the present devices (≈ 100 μm) was limited to the size large enough for us to obtain LMOKE signals.

To further demonstrate the reversible control of the magnetic state of the adjacent Co-layer by E-field, we have investigated how magnetic $H_C$ changes with the applied E-field in Co/BFO (200 nm) film heterostructure devices (200 μm × 200 μm). Figure 6(a) shows three representative LMOKE M(H) loops with magnetic field applied along the (110) direction of STO substrate in the as-grown state and after applying subsequent ± 10 V pulses. Clearly, the $H_C$ changes with different E-field pulses applied to the BFO layer. The $H_C$ is 4.8 mT in the as-grown state; it increases to 11 mT after a + 10 V pulse poling; and then it decreases to 6.8 mT after applying the second -10 V pulse. Based on a set of M(H) loops obtained after applying different voltage pulses, we have plotted the $H_C$ of the Co layer for the initial easy axis direction as a function of the applied E-field, as shown in Figure 6 (b). The black curve is the $H_C$ versus E-field in the first poling cycle where the device was poled with changing amplitude of the voltage pulses from the initial state to + 15 V, then to - 15 V, and finally back to 0 V where each voltage pulse was applied for 10 seconds. In the ascending branch, the $H_C$ increases slightly from the initial value of 4.8 mT after poling with pulses of + 2 V and + 5 V but rapidly jumps up to 11.0 mT at + 7.5 V.

The Hc value then remains at 11.0 mT while the amplitude of the voltage pulse further increases to as high as + 15 V. In the descending branch, the $H_C$ slightly decreases from 11.0 mT at the saturation state and then jumps down to 6.8 mT after a - 5 V voltage was applied. The Hc value then remains a constant for subsequently applied pulses with increasing negative voltage amplitude. We then preformed the same measurements on the same Co pad after the first poling cycle and measured the $H_C$ as a function of E-field for the second poling cycle in Fig. 6(b), denoted by the red curve. The second poling process shows two similar reversible $H_C$ states with the first poling process: the E-field dependent $H_C$ shows a hysteresis behavior with clear two reversible $H_C$ states (high $H_C$ and low $H_C$).

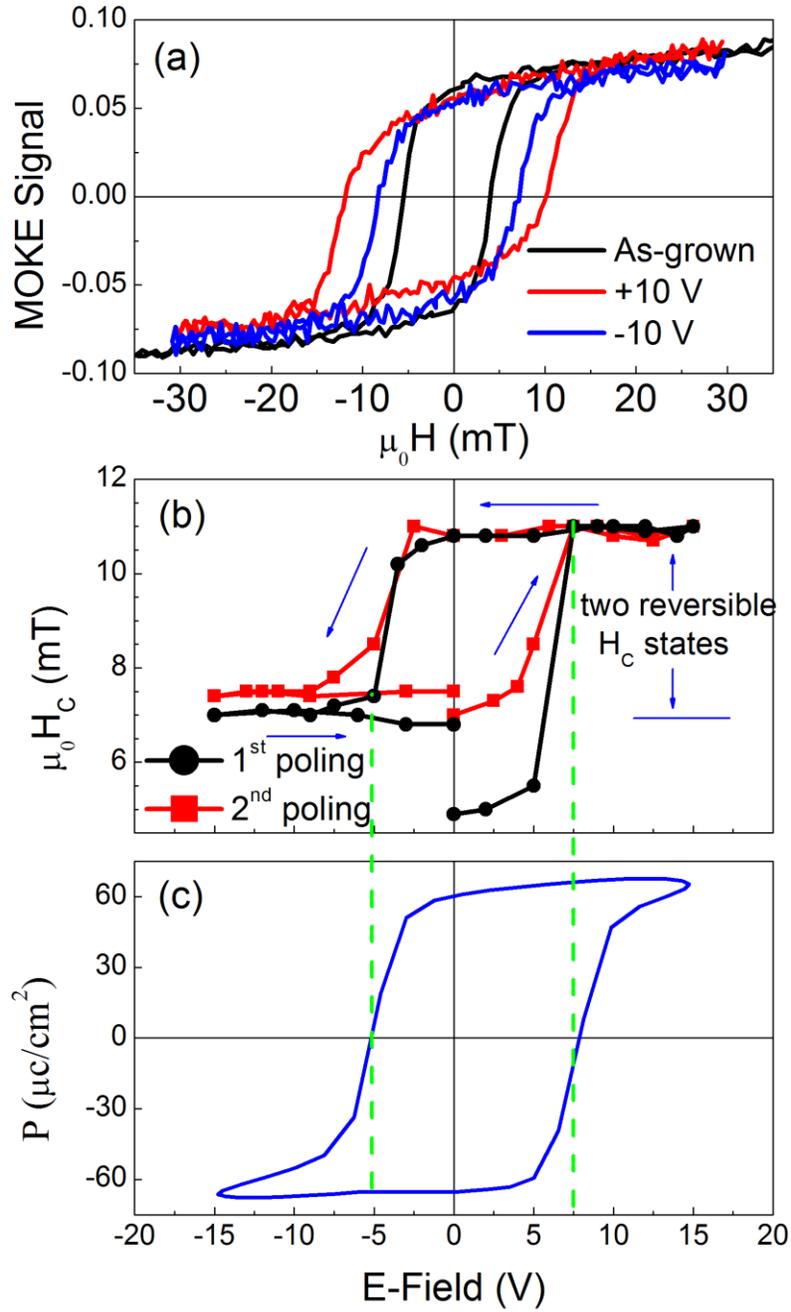

Figure 6 (a) LMOKE loops of Co (5 nm)/BFO (200 nm) at as-grown state, after + 10 V and - 10 V poling. The Co pad is $200\ \mu m \times 200\ \mu m$; (b) magnetic $H_C$ as a function of E-field value applied on BFO; (c) P-E (polarization – electric field) loop of 200 nm thick BFO film. The black and red curves in (b) represent the $H_C$ versus voltages at the first and second poling processes, respectively.

To confirm that the observed reversible $H_C$ states are directly correlated with ferroelectric states, we measured the polarization-electric field (P-E) loop of BFO film. As shown in Fig. 6(c), the P-E loop of 200 nm thick BFO film shows a saturation polarization of about 70 $\mu C/cm^2$ and the ferroelectric coercivity ($E_C$) is + 7.5 V and - 5 V in the ascending branch and descending branch, respectively. The switching of the $H_C$ states is coincident with FE polarization switching taking place at $E_C$ indicating that the observed two reversible $H_C$ states are directly associated with the switching of the FE polarization of BFO.

In the present work, the mechanism of the $H_C$-state switching is due to the reversible switching of the FE polarization in the BFO layer and the easy-axis direction in the adjacent Co layer. Because the $H_C$ states are tied to the reversible modulation of the magnetic anisotropy, the effect is expected to be robust and scalable. It provides an alternative pathway of reversible magnetic state modulation compared to the 180° switching of magnetization.

In summary, we have demonstrated reversible control of the magnetic state of a ferromagnetic Co layer by an E-field in Co/BFO film heterostructure devices fabricated on STO substrates. The LMOKE studies on the bilayer heterostructure suggest that the easy axis of Co is reversibly switched by 45° back and forth under the voltage pulses applied on the BFO. A new switching mechanism of the net vector of the exchange coupled Co-layer coupled to the canted moment of the AFM in multi-domain BFO is proposed to explain the 45° switching. These results demonstrate that the BFO devices fabricated on common STO substrates have the potential to serve as the basis for reversible E-field switchable spin valves with low energy consumption. The switching mechanism is a generic one since it arises naturally from the multi-domain state of BFO, and reversible modulation of magnetic anisotropy provides a possible path toward device scaling. Because epitaxial STO can be prepared on Si substrates[18-20], this opens the door for fabrication of reversible multiferroic spin valves on Si.

**Methods**

Methods and any associated references are available in the online version of the paper.

**Acknowledgement**

This work is supported by NIST (70NANB12H238, 70NANB15H261) and in part by C-SPIN, one of six centers of STARnet, a Semiconductor Research Corporation program, sponsored by MARCO and DARPA. We acknowledge S. Bowden, D. Pierce, and J. Unguris for the MOKE measurement facility at NIST.

**Author contributions**

I.T. and T.R.G. conceived the experiment. T.R.G. X.H.Z, S.M. M.M. P.J.C and A.V. made the samples and devices. X.H.Z. S.M and A.V. did the PFM measurements, and T.R.G. did the magnetic characterizations. R.W. and Z.Y. carried out the neutron diffraction experiments. T.R.G., I.T., W.R., and R.R. wrote the manuscript. All authors discussed and commented on the manuscript.

**Additional information**

Supplementary Information is available in the online version of the paper. Correspondence and requests for materials should be addressed to I.T.

**Competing financial interests**

The authors declare no competing financial interests.

**Methods**

Epitaxial BFO thin films used in this study were grown under 25 mT oxygen environment and at 590$^{\circ}$C using pulsed laser deposition on (001) oriented STO substrates with a pre-deposited 50 nm SrRuO$_3$ (SRO) layer, which serves as the bottom electrode. A 5 nm thick Co layer is deposited above the BFO layer by e-beam deposition which is patterned into arrays of square pads of different sizes using a lift-off process. A 5-nm Pd cap layer was deposited on the top of the Co layer to prevent its oxidization. The root mean square (RMS) surface roughness of our 100 nm thick BFO films is typically 2 nm measured over 5 μm ×5 μm area.

**Data availability**. The data that support the findings of this study are available from the corresponding author upon request.